# THE EFFECT OF CONTROLLED VIBRATIONS ON THE WIDTH BOUNDARY LAYERS DURING CRYSTAL GROWTH BY THE BRIDGMAN METHOD

A. I. Fedyushkin[*] and N. G. Bourago

Institute for Problems in Mechanics of RAS, Russia

## ABSTRACT

By using the finite element code ASTRA, the effect of vibrations on the melt flow in Bridgman crystal growth is investigated for the Earth and Space gravity conditions. It is found that the vibrations significantly decrease the width of interfacial boundary layers in Bridgman crystal growth with submerged vibrator.

The influence of the geometrical arrangement of the vibrator, the thermo-gravitational conditions, the frequency of vibration and the rotation effects is investigated. The averaged vibrational flow (AVF) parameters were calculated by means of post-processing the instant thermomechanical fields found in numerical integration of Navier-Stokes problem.

The structure of AVF depends on the physical properties of the melt and on the frequency/amplitude values. The frequency of vibrations (vibrational Reynolds number) influences essentially on the AVF structure. Numerical results show that the direction of AVF depends with frequency. It is shown that for normal and low gravity conditions the vibrations can decrease the width of boundary layers and increase the boundary temperature gradients. This can intensify heat and mass transfer near the solid-liquid interface and increase the rate of crystal growth.

## INTRODUCTION

The dopant distribution in the melt and consequently in the crystal depends on numerous conditions of the crystal growth. The study of the geometrical, thermal and dynamic issues such as forced and thermo-gravitational convection, rotation, magnetic field, vibration etc. is performed in many papers (see, for instance [1-11]). Besides scientific purposes of phenomenological study these investigations are practically targeted on searching the efficient control for the temperature and dopant distribution.

---

*Corresponding author, Alexey Fedyushkin fai@ipmnet.ru

In real conditions of vertical Bridgman crystal growth, the intensive melt convection always is present due to the horizontal heat fluxes in spite of the thermal scheme provides the stable temperature distribution and the cold melt-crystal interface is situated near the bottom of the crucible.

This convection essentially redistributes the dopant in the melt and does it not always in the desired manner. To decrease this convection the submerged baffles (or heaters) are in use [4,5]. In these modified crystal growth processes the convection flow is separated into two parts by submerged baffle/heater and has decreased convection intensity near the melt-crystal interface.

In Fig. 1 the vibrator is used as a submerged baffle. This case was under study in [4-6]. The comparison with the isothermal case indicated, that the influence of the convection on the temperature field is here negligible (especially under the vibrator). It was shown that weak mixing in the bottom part of crucible (near the solid-liquid interface) can cause the radial inhomogeneuity in the growing crystal even in microgravity environment. To level the dopant distribution in lower part of the crucible the various control tools are usably in use: the horizontal temperature gradient on the baffle surface [5] or on the baffle rotation [4].

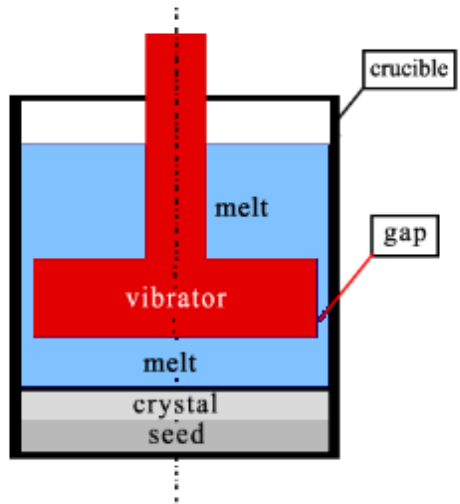
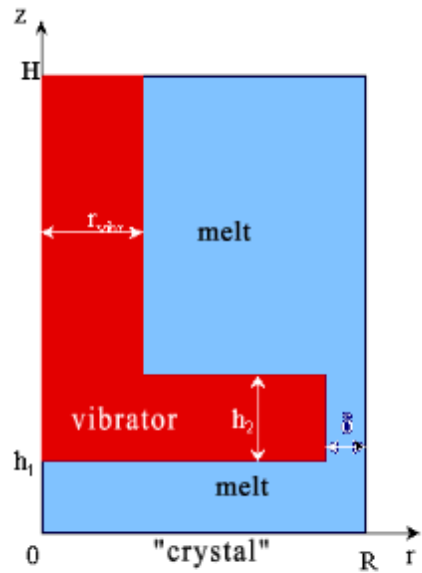

Fig. 1. Scheme of Bridgman crystal growth with submerged vibrator in the melt

Fig. 2. Scheme of calculation region

The vibrations can be used also as a control handle. In papers [7-11,15,16] the influence of vibrations on heat and mass transfer is studied theoretically and experimentally. In [9,10] the existence of two types of the averaged vibrational flow (AVF) was found: single vortex and double vortex flows which can be converted one into another depending on the frequency and the geometry of the vibrator.

Present study is devoted to numerical investigation of the influence of vibrations on the melt flow, the structure of AVF and the heat transfer in vertical directional Bridgman crystal growth under terrestrial and microgravity conditions. The calculated results indicate the strong influence of vibrations on the thickness of boundary layers.

The vibrations can decrease the boundary layers at the solid-liquid interface and they can represent the principal effect during the crystal growth process. It is found that by using the vibrations it is possible to vary the value of temperature gradient at the solid-liquid interface, i.e., to vary the kinetics and the rate of the crystal growth. This effect takes place in both terrestrial and microgravity conditions and is dominant for considered configurations. It even prevails on the thermal convection. The effect of Prandtl numbers on the AVF is also studied.

## 1. PROBLEM STATEMENT

The modelling conducted on the basis of numerical solution of unsteady 2D Navier-Stokes equations for incompressible axisymmetrical flow. The finite element method and the computer code "ASTRA" [12,13] used. This code allows to solve 2D and 3D nonstationary Navier-Stokes problems in complex geometry regions.

The AVF was defined by averaging the history of instantaneous unsteady Navier-Stokes solution directly, calculated step-by-step in time. It permits to observe the evolution of the AVF in time and space and its dependence on problem parameters.

The necessity of big CPU time resources was one of the major difficulties of the numerical modelling. For instance the single run can require several tens hours of CPU time on Pentium III 500 MHz.

The principal geometrical scheme of Bridgman crystal growth with submerged vibrator is depicted in Fig. 2. In the mathematical model the following assumptions were used:

axial symmetry, crystal growth rate and thermal conditions are permanent during the whole process, melt-crystal interface is flat. The small amplitude vibrations are under consideration and therefore the vibrator displacements are negligible while the vibrator velocity is a predefined harmonic function $v = A\omega \sin(\omega t)$, where A and $\omega$ - an amplitude and a frequency of vibrations.

Unsteady 2D asisymmetrical Navier-Stokes-Boussinesq equations for incompressible fluid read

$$\frac{\partial u}{\partial r} + \alpha \frac{u}{r} + \frac{\partial w}{\partial z} = 0 \tag{1}$$

$$\frac{du}{dt} - \alpha \frac{v^2}{r} = -\frac{1}{\rho}\frac{\partial p}{\partial r} + \frac{1}{r^\alpha}\frac{\partial}{\partial r}\left(r^\alpha v \frac{\partial u}{\partial r}\right) + \frac{\partial}{\partial z}\left(v \frac{\partial u}{\partial z}\right) - \alpha v \frac{u}{r^2} \tag{2}$$

$$\frac{dw}{dt} = -\frac{1}{\rho}\frac{\partial p}{\partial z} + \frac{1}{r^\alpha}\frac{\partial}{\partial r}\left(r^\alpha v \frac{\partial w}{\partial r}\right) + \frac{\partial}{\partial z}\left(v \frac{\partial w}{\partial z}\right) + g\beta_T(T - T_0) \tag{3}$$

$$\frac{d\rho c_p T}{dt} = \frac{1}{r^\alpha}\frac{\partial}{\partial r}\left(r^\alpha \lambda \frac{\partial T}{\partial r}\right) + \frac{\partial}{\partial z}\left(\lambda \frac{\partial T}{\partial z}\right) \tag{4}$$

where r and z - radial and axial coordinates, t - time, u and w - velocity components in r and z directions, accordingly, T - a temperature, p - a pressure, $\rho$ - a density, g - a gravity acceleration, $\beta_T$ - a thermal expansion factor, $v$ - a kinematic viscosity, $\lambda$ - a heat conduction, $c_p$ - a heat capacity at constant pressure, $\alpha = 1$ - in case of axisymmetrical flow and $\alpha = 0$ - in case of flat flow. The boundary condition read:

axis of symmetry ($r = 0$): $u = 0$, $\frac{\partial w}{\partial r} = 0$, $\frac{\partial T}{\partial r} = 0$; \hfill (5)

solid-liquid interface ($z = 0$): $u = 0$, $w = -W_S$, $T = T_m$; \hfill (6)

side wall of crucible ($r = R$): $u = 0$, $w = 0$, $\frac{\partial T}{\partial r} = 0$ ($0 < z < h$), $T = T_h$ ($h < z < H$); \hfill (7)

vibrator: $u = 0$, $w = A\omega \sin(\omega t)$, $\frac{\partial T}{\partial n} = 0$ \hfill (8)

upper boundary ($z = H$): $u = 0$, $\frac{\partial w}{\partial z} = 0$, $T = T_h$ \hfill (9)

Initial conditions read: (t=0): $u = 0$, $w = 0$, $T = T_m$. \hfill (10)

$W_S$ - crystal growth rate, $\Delta T$ - temperature scale, **n** - normal unit vector.

Initial temperature distribution $T_m$ (r, z) was calculated by solving steady temperature equation. Crystal growth rate reads: $W_S = 0.3$ sm/h. The problem is characterized by the following similarity numbers:

Reynolds number $Re = W_S R / \nu$, vibrational Reynolds number: $Re_{vibr} = A\omega R / \nu$, Grashof number: $Gr = g\beta\Delta T R^3 / \nu^2$, Rayleigh number: $Ra = GrPr$, Prandtl number: $Pr = \nu \rho c_p / \lambda$.

The values of these numbers were: $Re_{vibr}$ and $Re < 10$ ; $Gr = 0 \div 2.1 \cdot 10^6$, $Pr = 0.01 \div 54$.

## 2. SOLUTION TECHNIQUE

The finite element method is used. The approximations are based on linear triangular elements in space and on implicit finite difference scheme in time. Auxiliary boundary value problems for stream function, vorticity and temperature were solved by using conjugate gradient method with preconditioning without matrix operations [10]. Convective and diffusive terms were approximated symmetrically, for stabilization the additional dissipative diffusion terms were added to each convection-diffusion equation.

The code was tested by solving known benchmark problems [4,5] and the results of tests for thermal convection in quadratic region [7] indicated good accuracy of the code [14, 15].

The AVF flow characteristics were evaluated by using time averaging formula:

$$F_{average} = \frac{1}{t}\int_0^t F \, dt \quad (11)$$

where t is time, F - some of the flow parameters.

The calculations were made using the time step value so that its value was 1/100-1/10 of one period of vibrations (depending on the frequency of vibrations). All presented here results correspond to the time when the averaging vibrational flow (AVF) becomes quasistationary.

## 3. NUMERICAL RESULTS

Below the results of the parametric calculations are presented for variety of geometrical configurations of crucibles with submerged vibrators. Considered regions had the following sizes: R=1.6; H=3, 3.2; $r_{vibr}$=0.4; $h_1$=0.8; $h_2$=0.8; $\delta = 0.1$ (cm);

where R – a radius of the ampule, H – a height of the ampule, $h_1$ – a distance between the vibrator and the solid-liquid interface, $h_2$ - a vibrator thickness (a distance between its lower and upper surfaces), $\delta$ - a gap (a distance between the vibrator and side wall of the crucible. The following variants with dimensional values of A=50µм and 100µм, $f = \omega/2\pi$=0-100 (Hz) are calculated.

### 3.1 Averaged vibrational flow (AVF)

Vibrating baffle forces the melt flow which regarding to time scale can be represented by two flows of different nature with different flow characteristics. First representation is the instantaneous vibrational convective melt flow, which is controlled by the vibrator frequency. This flow can not be detected visually in experiments at rather high vibrator frequencies f=10-100 Hz. Second representation is the AVF, which has significantly decreased velocities and soon becomes quasistationary. This flow can be observed as particle tracks in the experiments with transparent liquids. It is not gravity sensitive and exists as well in case of microgravity and under isothermal conditions. Under non-isothermal conditions when the vibrational flow and thermal convection flow are present simultaneously, the integrated flow is finally observed.

The influence of vibrations (A=100 µm and f=50 Hz) for microgravity conditions (without thermal convection) is demonstrated in Fig.3 for $NaNO_3$ melt ($\nu$=0.015[$sm^2$/sec], a=0.0158 [$sm^2$/sec], Pr=9.49). The isolines of stream function are depicted there for instantaneous flow (Fig.3b) and for the AVF (Fig.3c) for the value of time t=20 sec. The AVF is present in the areas above and under vibrator so as if it is separated by the vibrator into two parts, which have single-vortex structures. The AVF has the clockwise direction under the vibrator and piecewise direction above. Besides in Fig 3 the graphs of stream function (Fig.3a) and averaged stream function (Fig.3b) versus time are depicted. The results of calculations indicate that the stationary regime of the AVF (for used input parameters) appears after 15-20 seconds (Fig.3c).

## 3.2 Influence of vibrations on temperature boundary layer near the solid-liquid interface

Influence of vibrations on temperature distribution is shown in Fig.4-5. In Fig.4 - for terrestrial environment, and in Fig.5 - for microgravity. In Fig. 4 the results without vibrations (f=0, Fig.4a, c, e) and with vibrations (A=100 μm и f=50 Hz, Fig.4 b, d, and f) can be seen for the following similarity numbers: $Gr = 2.1 \cdot 10^6$, $Pr = 5.43$, $H/R = 3.2$. Fig.4a and Fig. 4b indicate the isolines of stream function, Fig.4c and 4d - isotherms and Fig.4e и 4f - vertical profiles of the temperature (r=0.75). Under influence of vibrations the large scale AVF is formed. The structure of the AVF in this case has simple arrangement: it is single vortex above the vibrator and double-vortex under (see Fig. 4b). Above the vibrator the vortex rotates in counterclockwise direction. Under the vibrator there is the dominant vortex of low intensity rotating in clockwise direction and some secondary vortex of bigger size rotating in counterclockwise direction. Negative values of stream function correspond to the flow rotating in clockwise direction while positive values to the opposite flow.

The vibrational effect essentially increases the temperature gradient at the solid-liquid interface. This can be seen from comparison of the results in Fig. 4c, e and in Fig. 4d, f.

It should be told that the fact of narrowing the temperature boundary layer results not only from the AVF itself but from the type of instantaneous flow as well. The type of vibrations plays the principal role. Undoubtedly the instantaneous flow forms the boundary layers. The intensity of instantaneous flow more than 10 times higher than the intensity of the AVF.

The influences of thermal convection on the variation of vertical temperature gradient at the solid-liquid interface (in the area under the vibrator) is practically absent. This can be seen from comparison between the results for terrestrial conditions (Fig.4) and for microgravity conditions (Fig.5). So, in this case the vibrations become a dominant effect forming the temperature distribution and temperature boundary layers at the solid-liquid interface.

The influence of vibrations on the boundary layers depends on the amplitude, the frequency and, also on the Prandtl number of the melt. The calculations indicate, that for small value of Prandtl number Pr=0.01, f=50Hz, A=100μm due to small viscosity and big thermal conductivity the influence of vibrations on the temperature boundary layer thickness is negligible. The influence of vibrations on the temperature boundary layer in melts with small Prandtl numbers can become essential with increase of the amplitude and the frequency is increased (with increase of vibrational Reynolds number $Re_{vibr}$).

Instantaneous flow

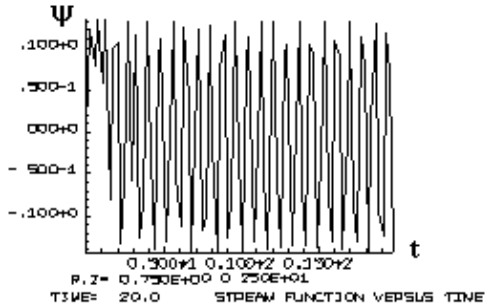

a) Character of change of a stream function on time
(in a point with coordinates r=0.75cm, z=2.5cm)

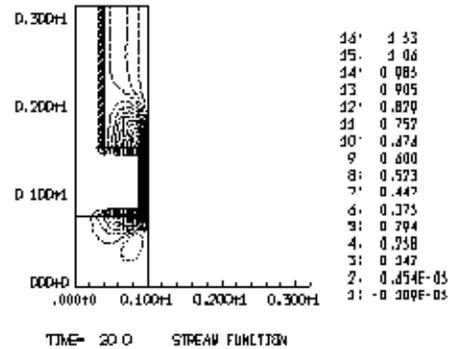

b) Isolines of stream function on time
(in an instant of time 20sec)

Average vibration flow (AVF)

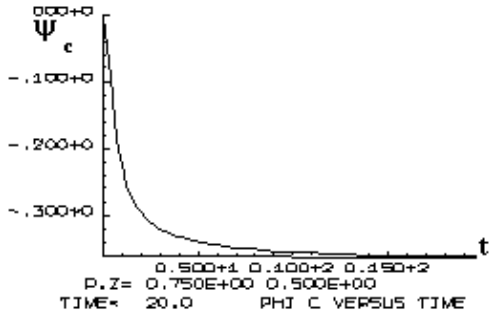

c) Change of an average stream function on time

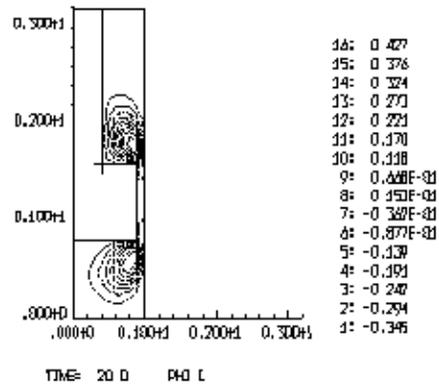

d) Isolines of stream function of AVF
(in an instant of time 20sec)

Fig.3 Vibrational flow in the melt of $NaNO_3$
(isothermal case, $g/g_0=0$, A=100 μm, f =50 Hz, R=1 cm, H=3,2 cm)

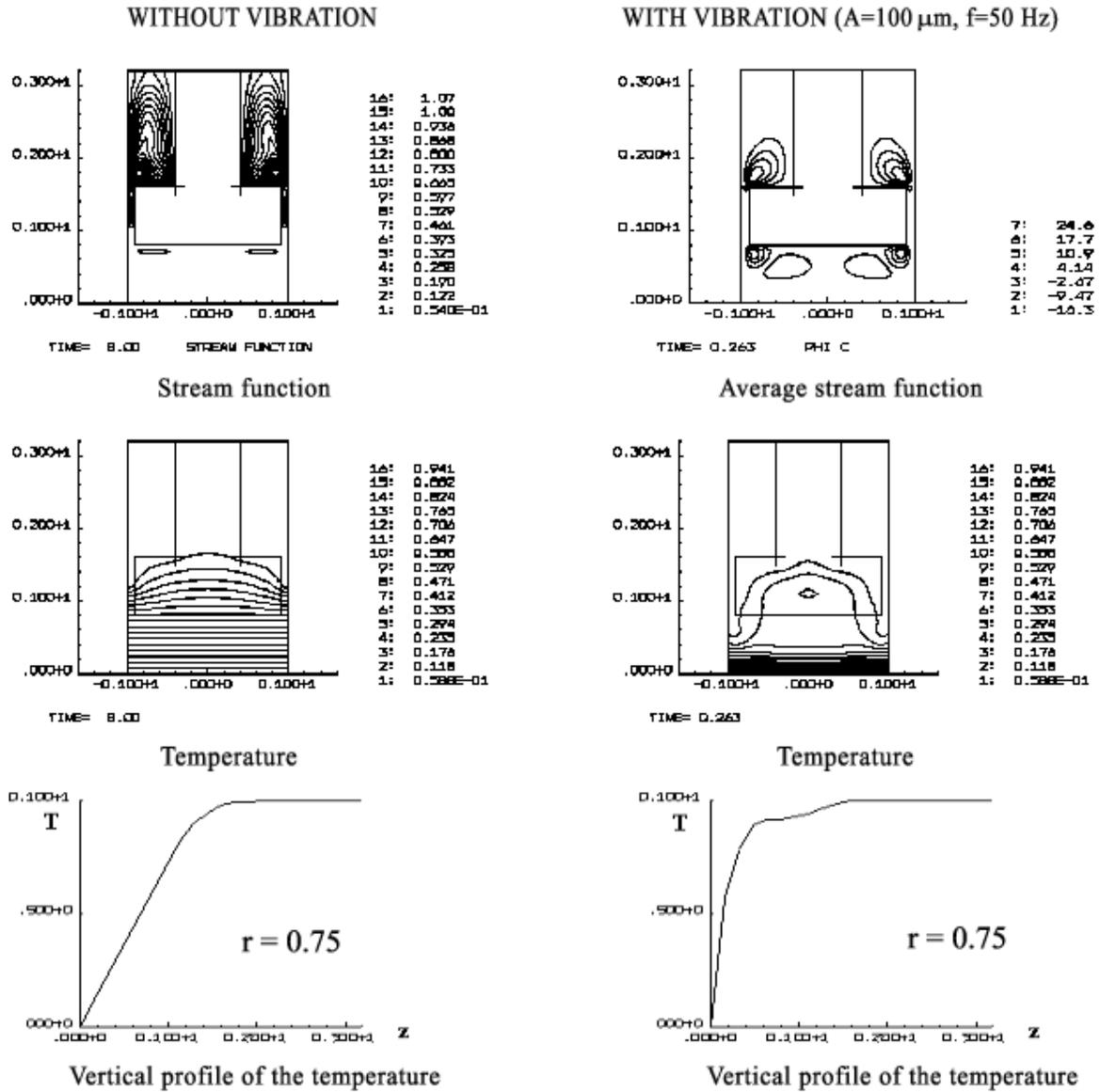

Fig.4 Influence of vibrations on a temperature boundary layer at front of a crystallization ( Gr=2.1 $10^6$, Pr=5.43, H/R=3.2)

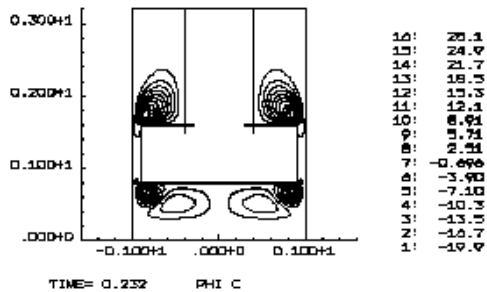
Average stream function

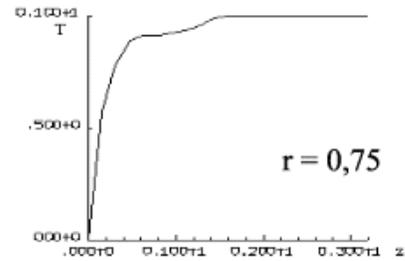
Vertical profile of the temperature

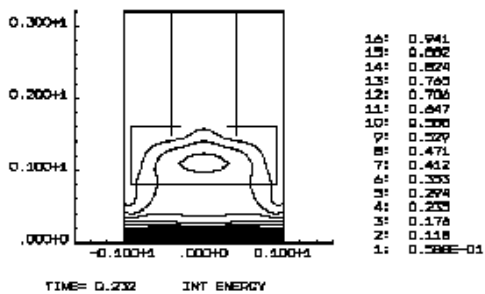
Temperature

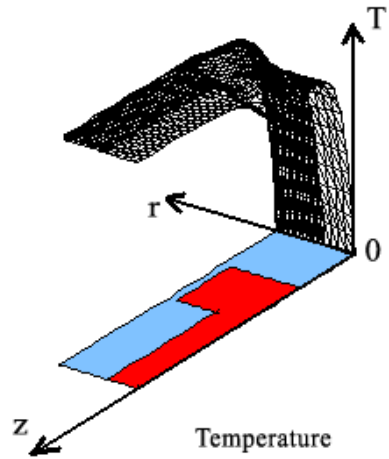
Temperature

Fig.5 Influence of vibrations on a temperature boundary layer at front of a crystallization in space environment
(Gr=0, $g/g_0$=0 Pr=5.43, H/R=3.2, A=100 μm, f=50 Hz)

### 3.3 The AVF structure under various values of Grashof and Prandtl numbers

The isolines of the AVF stream function under various Grashof and Prandtl numbers for f=50 Hz and A=100 μm are presented in Fig.6: a) Gr=0, Pr=5.43; b) Gr=2.1 $10^5$, Pr=0.18; c) Gr=2.1 $10^5$, Pr=5.43; d) Gr=2.1 $10^5$, Pr=54.3; e) Gr=2.1 $10^6$, Pr=5.43; f) Gr=2.1 $10^6$, Pr=0.01. Under variation of the Prandtl number the AVF structure near the solid-liquid interface can be changed from single vortex to double vortex. The AVF structures under the vibrator for big Prandtl numbers can become layered. The layered structure flows are observed in vertically stratified fluids (under big Prandtl numbers these structures appear when the vertical temperature stratification in homogeneous fluid takes place). It should be mentioned that the results, which were not included into present paper, indicated that the direction of the AVF can be changed by variation of frequency of the vibration [16-21].

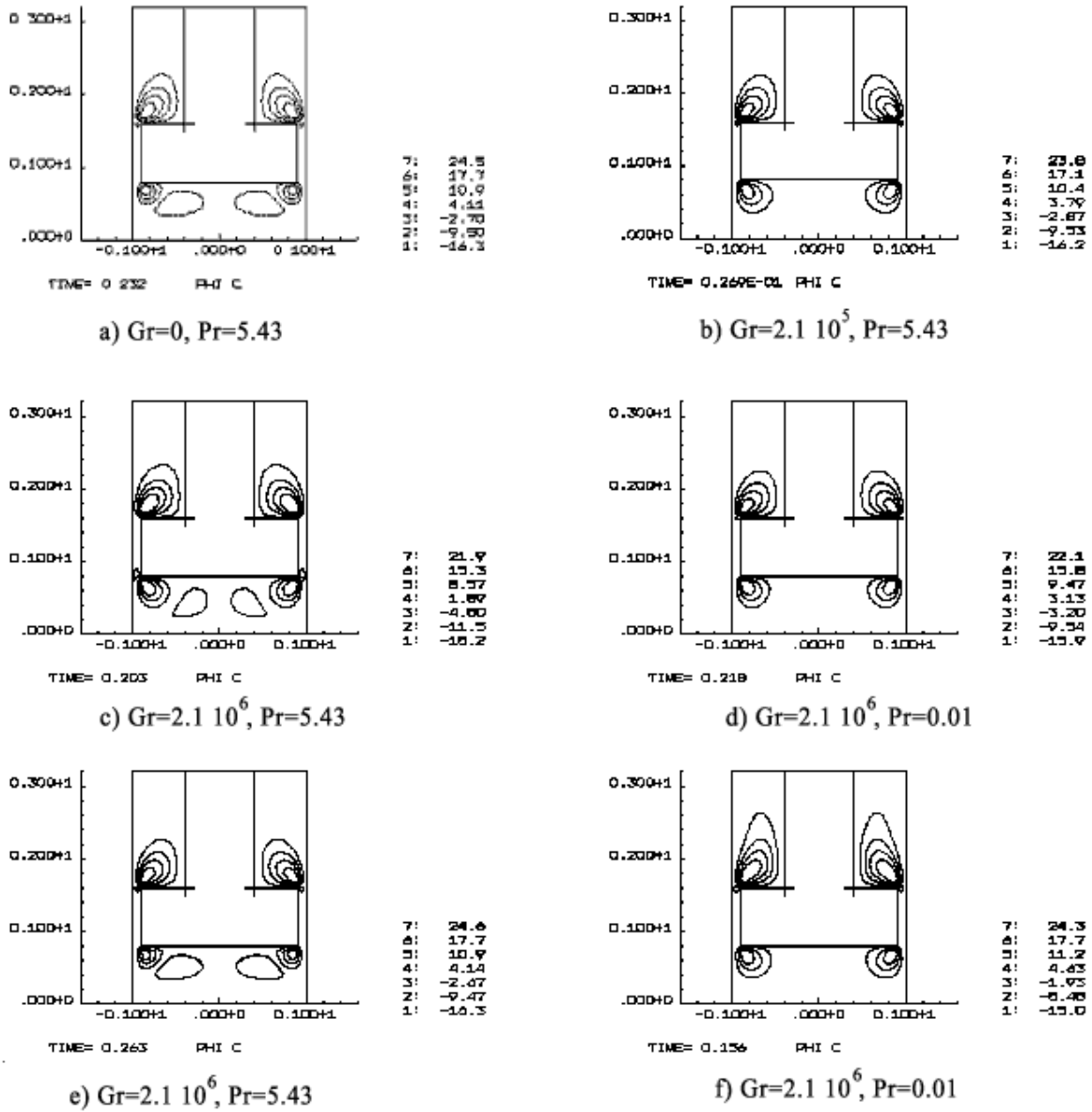

Fig.6 Average vibration flow for different Grashoff and Prandtl numbers
(A= 100 μm, f=50 Hz)

## CONCLUSIONS

For various values of similarity numbers ($Re_{vibr}$, Pr, Gr) by means of the numerical Navier-Stokes modelling the character of the AVF is investigated. The influence of the vibrations on the temperature boundary layer is highlighted for variety of Prandtl numbers.

It is found that the vibrations can decrease the boundary layer thickness at the solid-liquid interface. It can take the principal meaning for improvement of crystal growth technologies due to the possibility to use the vibrations for the control of temperature gradients at solid-liquid interface, i.e. for the control of the kinetics and rate of crystal growth. This effect takes place in both terrestrial and microgravity conditions and for considered configurations the vibrations play the dominant role compared to natural convection.

The results of parametric calculations indicated, that the vibrations could appear a simple, cheap and effective governing factor (in comparison with microgravity, heat transfer and magnetic field), which controls the hydrodynamics, heat and mass transfer, dopant distribution and crystal growth kinetics.

## REFERENCIES